\documentclass[aps,twocolumn,showpacs]{revtex4}
\usepackage{graphicx}
\newcommand{\bqa}{\begin{eqnarray}}
\newcommand{\eqa}{\end{eqnarray}}
\newcommand{\pslash}{\slash\hspace{-0.55em}}
\newcommand{\as}{\alpha_{\mathrm{s}}}

\begin{document}

\title{Infrared divergences of $B$ meson exclusive decays to
P-wave charmonia in QCD factorization and nonrelativistic QCD}
\author{Zhong-Zhi Song, Ce Meng, Ying-Jia Gao}
\affiliation{Department of Physics, Peking University,
 Beijing 100871, People's Republic of China}
\author{Kuang-Ta Chao}
\affiliation{China Center of Advanced Science and Technology
(World Laboratory), Beijing 100080, People's Republic of China;\\
Department of Physics, Peking University, Beijing 100871, People's
Republic of China}
\date{\today}
\begin{abstract}
In the framework of QCD factorization, we study the $B$ meson
exclusive decays $B\rightarrow \chi_{cJ}K$ where the spin-triplet
P-wave charmonium states $\chi_{cJ}(J=0,1,2)$ are described by the
color-singlet non-relativistic wave functions. We find that for
these decays (except $\chi_{c1}$) there are infrared divergences
arising from nonfactorizable vertex corrections as well as
logarithmic end-point singularities arising from nonfactorizable
spectator interactions at leading-twist order. The infrared
divergences due to vertex corrections will explicitly break down
QCD factorization within the color-singlet model for charmonia.
Unlike in the inclusive decays where the higher Fock states with
color-octet $c \bar c$ pair and soft gluon can make contributions
to remove the infrared divergences, their contributions can not be
accommodated in the exclusive two body decays.  As a result, the
infrared divergences encountered in exclusive processes involving
charmonia may raise a new question to the QCD factorization and
NRQCD factorization in B exclusive decays.
\end{abstract}
 \pacs{13.25.Hw; 14.40.Gx; 12.38.Bx  \hfill}
\maketitle
$B$ meson decays, especially exclusive nonleptonic
decays, provide an opportunity to determine the parameters of the
Cabibbo-Kobayashi-Maskawa (CKM) matrix, to explore CP violation
and to observe new physics effects. However, although the
underlying weak decay of the $b$ quark is simple, quantitative
understanding of nonleptonic $B$-meson decays is difficult due to
the complicated strong-interaction effects.

Exclusive $B$-meson decays to final states containing charmonia
are especially important due to three reasons. First, those decays
e.g. $B\to\!J/\psi K$ are regarded as the golden channels for CP
studies due to their clean experimental signatures and
straightforward theoretical interpretations. Secondly, those
decays may provide useful information towards the understanding of
color suppressed decays. Thirdly, those decays involve two energy
scales, the beauty quark mass $m_b$ and charm quark mass $m_c$,
and therefore are more subtle in theoretical studies. Since
experimentally CLEO, BaBar and Belle Collaborations have provided
many measurements on the subject of $B$ exclusive decays to
charmonia \cite{ex} such as $J/\psi, \psi', \eta_c, \eta_c',
\chi_{c0}$, and $ \chi_{c1}$, theoretical studies on these issues
become necessary.

We will start with the QCD factorization approach proposed by
Beneke et al.\cite{BBNS}. It is argued that because the size of
charmonium is small$(\sim \!\!1/{\as {m_\psi}})$ and its overlap
with the $(B, K)$ system is negligible, the same QCD factorization
method as for $B\to\!\pi\pi$ can be used for $B \to\!J/\psi K$
decay. This small size argument for the applicability of QCD
factorization to $B$ exclusive decays to charmonia needs to be
tested for specific decay channels. Indeed, explicit calculations
for $B \to\!J/\psi K$ within the QCD factorization
approach\cite{ch} showed that the nonfactorizable vertex
contribution is infrared safe and the spectator contribution is
perturbatively calculable at twist-2 order though the theoretical
branching ratio is much smaller than the experimental data.
Studies on $B \to\! \eta_c K$ decay\cite{song1} also confirmed the
qualitative applicability of the small size argument while the
quantitative underestimates compared with data were very similar
to that encountered in $B \to\! J/\psi K$. However, further
studies on $B \to\! \chi_{c0} K$ \cite{song2} challenged the
applicability of QCD factorization for charmonia because of the
appearance of non-vanishing infrared divergences in this decay.

In previous studies for exclusive $B$ decays to charmonia within
the QCD factorization approach, the light-cone wave function
description of the charmonia was adopted. This is fairly
appropriate for the S-wave charmonia since the relative momentum
$q$ between charmed and anti-charmed quarks can be neglected in
the lowest order approximation. However, for P-wave charmonia, $q$
can not be neglected even in the leading-order. Although the
light-cone wave function description of  charmonia can provide us
with some essential features of the problems involved in these
decays, it can not give a complete analysis. The more appropriate
method is to study charmonia within the non-relativistic
bound-state picture or equivalently within the nonrelativistic QCD
(NRQCD) framework\cite{BBL1}. In this letter, we report a complete
analysis of the infrared unsafety encountered in $B$ exclusive
decays to P-wave charmonium states, including $\chi_{c2}$, which
was not involved in previous studies, by using the color-singlet
non-relativistic wave functions to describe the charmonia.

To proceed, we first give a brief review of the QCD factorization
approach. The general idea of the QCD factorization is that in the
heavy quark limit $m_b\! \gg \Lambda_{\mathrm{QCD}}$, the
transition matrix elements of operators in the hadronic decay $B
\to\! M_1M_2$ assumes a simple form at the leading order in
$1/m_b$\cite{BBNS}:
 \bqa \label{qcdf}
\langle M_1M_2|{\cal O}_i|B\rangle = F^{BM_1}(m_2^2)\int^1_0
dx\,T^I(x)\phi_{M_2}(x) \nonumber \\ +\int^1_0 d\xi \,dx\,dy
\,T^{II}(\xi,x,y)\phi_B(\xi)\phi_{M_1}(y)\phi_{M_2}(x),
 \eqa
where $M_1$ is the recoiled meson and $M_2$ is the emitted meson
which is a light meson or a quarkonium. $F^{BM_1}$ is the $B \to\!
M_1$ transition form factor, $\phi_M$ is the light-cone
distribution amplitude and $T^{I,II}$ are perturbatively
calculable hard scattering kernels. If we neglect strong
interaction corrections, formula.(\ref{qcdf}) reproduces the
result of naive factorization. However, hard gluon exchange
between $M_2$ and $BM_1$ system implies a nontrivial convolution
of hard scattering kernels $T^{I,II}$ with the distribution
amplitude $\phi_{M_2}$.

In the non-relativistic bound-state picture, instead of the
light-cone amplitude distribution, charmonium can be described by
the color-singlet non-relativistic wave function (the role of the
higher Fock states with color-octet $c \bar c$ pair will be
considered later on). Let $p_\mu$ be the total 4-momentum of the
charmonium and $q_\mu$ be the relative 4-momentum between $c$ and
$\bar c$ quarks. For P-wave charmonium, because the wave function
at the origin $\mathcal{R}_1(0)\!\!=\!\!0$, which corresponds to
the zeroth order in $q$, we must expand the amplitude to first
order in $q$. Thus we have (see, e.g., \cite{kuhn})
 \bqa
 \label{amp}
\mathcal{M}(B\to\!\! {}^{2S+1}\!P_J(c\bar
c))\!=\!\!\!\!\sum_{L_z,S_z}\!\langle 1L_z;SS_z|JJ_z\rangle
 \!\int\!\!\frac{\mathrm{{d}}^4 q}{(2 \pi)^3}q_\alpha \nonumber\\
 \times \delta\!(q^0\!\!-\!\!\frac{|\vec{q}|^2}{M}\!)\psi_{1M}^\star\!(q)
 \mathrm{Tr}[\mathcal{O}^\alpha\!(0)P_{SS_z}\!(p,\!0)
\!+\!\mathcal{O}\!(0)P^\alpha_{SS_z}\!(p,\!0)],
 \eqa
where $\mathcal{O}(q)$ represents the rest of the decay matrix
element and the spin projection operators $P_{SS_z}(p,q)$ which is
constructed in terms of quark and anti-quark spinors as
 \bqa P_{SS_z}(p,q)\!=\!\!\sqrt{\!\frac{3}{m}}\!\sum_{s_1,s_2}\!\!v(\frac{p}{2}\!-\!q,\!s_2)
 \bar u(\!\frac{p}{2}\!+\!q,\!s_1) \!\langle s_1;\!s_2|SS_z\!\rangle,
 \eqa
 and
 \bqa
\mathcal{O}^\alpha(0)\!&=&\!\frac{\partial
\mathcal{O}(q)}{\partial
q_\alpha}|_{q=0},\\
 P^\alpha_{SS_z}(p,0)\!&=&\!\frac{\partial P_{SS_z}(p,q)}{\partial q_\alpha} |_{q=0}.
  \eqa

After $q^0$ is integrated out, the integral in Eq.(\ref{amp}) is
proportional to the derivative of the P-wave wave function at the
origin by
 \bqa
\int\!\frac{\mathrm{{d}}^3 q}{(2 \pi)^3}q^\alpha \psi_{1M}^\ast(q)
=i\varepsilon^{\ast\alpha}(L_z)\sqrt{\frac{3}{4\pi}}
\mathcal{R}^{'}_1(0), \eqa where $\varepsilon^\alpha(L_z)$ is the
polarization vector of an angular momentum one system and the
value of $\mathcal{R}^{'}_1(0)$ for charmonia can be found in e.g.
Ref.\cite{quig}

The spin projection operators $\!P_{SS_z}(p,0)\!$ and
$\!P^\alpha_{SS_z}(p,0)\!$ can be written as\cite{kuhn}
 \bqa
P_{1S_z}(p,0)&\!\!=\!\!&\sqrt{\frac{3}{4 M}}\pslash
\varepsilon^\ast(S_z)(\pslash p+M),\\
P^\alpha_{1S_z}(p,0)&\!\!=\!\!\!&\sqrt{\frac{3}{4 M^3}}[\pslash
\varepsilon^\ast(S_z)(\pslash
p+\!M)\gamma^\alpha\!+\!\gamma^\alpha \pslash
\varepsilon^\ast(S_z)(\pslash p+\!M)]\nonumber
 \eqa
where we have made use of the non-relativistic approximation for
the charmonium mass $M\simeq 2 m$. Here $m$ is the charmed quark
mass.

In the calculation we need following polarization relations for
the ${}^3P_J$ states
 \bqa\label{pol}
 \sum_{L_Z
S_Z}\!\!\epsilon^{\ast\alpha}\!(\!L_z\!)\epsilon^{\ast\beta}\!(\!S_z\!)
\langle
1L_z;\!1S_z|00\rangle\!&=&\!\frac{1}{\sqrt{3}}(-\!g^{\alpha\beta}+\frac{p^\alpha
p^\beta}{M^2}\!),\nonumber\\
\sum_{L_Z
S_Z}\!\!\epsilon^{\ast\alpha}\!(\!L_z\!)\epsilon^{\ast\beta}\!(\!S_z\!)
\langle 1L_z;\!1S_z|1J_z\rangle\!&=&\! \frac{-i
\epsilon^{\alpha\beta\lambda\kappa}p_\kappa\epsilon_\lambda^\ast\!(\!J_z\!)}
{\sqrt{2}M},\nonumber\\
 \sum_{L_Z
S_Z}\!\!\epsilon^{\ast\alpha}\!(\!L_z\!)\epsilon^{\ast\beta}\!(\!S_z\!)
\langle
1L_z;\!1S_z|2J_z\rangle\!&=&\!\epsilon^{\ast\alpha\beta}(J_z),\,\,\,
 \eqa
where $\epsilon_\lambda(J_z)$ is the usual spin-1 polarization
vector and the polarization tensor $\epsilon^{\alpha\beta}(J_z)$
is that appropriate for a spin-2 system which is symmetric under
the exchange $\alpha\leftrightarrow\!\beta$.

The effective Hamiltonian relevant for $B \to\! \chi_{cJ}K$ is
written as\cite{BBL}
 \bqa
\mathcal{H}_{\mathrm{eff}}\!\!=\!\!\frac{G_F}{\sqrt{2}} \Bigl(\!
V_{cb} V_{cs}^*\!(C_1 {\cal O}_1\!+C_2 {\cal O}_2 )\!-V_{tb}
V_{ts}^* \sum_{i=3}^{6} C_i {\cal O}_i \!\Bigr),
 \eqa
where $G_F$ is the Fermi constant, $C_i$ are the Wilson
coefficients and $V_{q_1q_2}$ are the CKM matrix elements. We do
not include the effects of the electroweak penguin operators since
they are numerically small. Here the relevant operators ${\cal
O}_i$ are given by
 \bqa
{\cal O}_1\!&=&\!(\overline{s}_{\alpha} b_{\beta})_{V-A} \cdot
(\overline{c}_{\beta} c_{\alpha})_{V-A},
 \nonumber\\
 {\cal O}_2\!&=&\!(\overline{s}_{\alpha} b_{\alpha})_{V-A} \cdot
(\overline{c}_{\beta} c_{\beta})_{V-A},
 \nonumber\\
{\cal O}_{3(5)}\!&=&\!(\overline{s}_{\alpha} b_{\alpha})_{V-A}
\cdot \sum_q (\overline{q}_{\beta} q_{\beta})_{V-A(V+A)},
\\
{\cal O}_{4(6)}\!&=&\!(\overline{s}_{\alpha} b_{\beta})_{V-A}
\cdot \sum_q (\overline{q}_{\beta} q_{\alpha})_{V-A(V+A)},
\nonumber
 \eqa
where $\alpha,~\beta$ are color indices and the sum over $q$ runs
over $u, d, s, c$ and $b$. Here $(\bar q_1 q_2)_{V\pm A}=\bar
q_1\gamma_\mu (1\pm\gamma_5) q_2 $.

\begin{figure}[t]
\vspace{-2.0cm}
 \hspace{-1.5cm}
\includegraphics[width=10cm,height=11cm]{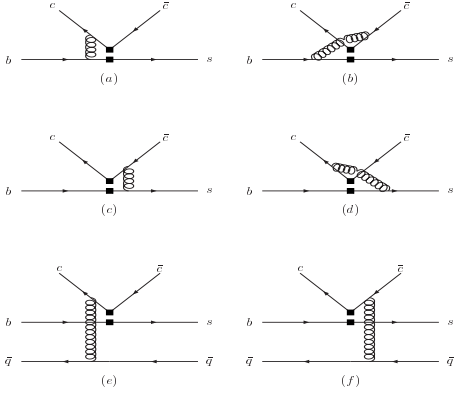}
\vspace{-3.5cm}
\caption{ Feynman diagrams for vertex and
spectator corrections to $B \to\! \chi_{cJ} K$.} \label{fvs}
\end{figure}

According to \cite{BBNS} all nonfactorizable corrections are due
to Fig.\ref{fvs}, and other corrections are factorized into the
physical form factors and meson wave functions. Taking
nonfactorizable corrections in Fig.\ref{fvs} into account, the
decay amplitude for $B\to\! \chi_{cJ}K(J=0,2)$ in QCD
factorization is written compactly as
 \bqa\label{amp2}
  i\mathcal{M} =\frac{G_F}{\sqrt{2}}\Bigl[V_{cb}
V_{cs}^* C_1-V_{tb} V_{ts}^* (C_4 + C_6) \Bigr]\times A,
 \eqa
where the coefficients $A$ are given by  \bqa\label{a} A=\frac{i 6
\mathcal{R}^{'}_1} {\sqrt{\pi
M}}\cdot\frac{\alpha_s}{4\pi}\frac{C_F}{N_c} \Bigl(F_1\cdot f_I +
\frac{4\pi^2 f_K f_B}{N_c}  \cdot f_{II} \Bigr).
 \eqa
Here $N_c$ is the number of colors, $C_F=(N_c^2-1)/(2 N_c)$, and
$F_1$ is the $B \to K$ form factor (see Eq.(\ref{vmu2}) below).
The function $f_I$ is calculated from the four vertex correction
diagrams (a, b, c, d) and $f_{II}$ is calculated from the two
spectator correction diagrams (e, f) in Fig.\ref{fvs}.

The form factors for $B \to\! K$ are given as
  \bqa \label{vmu2}
&&\langle K(p_K) | \overline{s} \gamma_{\mu} b| B(p_B)\rangle= \\
&&F_1 (p^2)(p_B +p_K)_{\mu}+ [F_0 (p^2)-F_1
(p^2)]\frac{m_B^2-m_K^2}{p^2} p_{\mu},\nonumber
  \eqa
where $p= p_B -p_K$ is the momentum of charmonium  with $p^2=M^2$,
and $m_B, m_K$ are respectively the masses of $B$, $K$ mesons. We
will neglect the kaon mass for simplicity. We can use the ratio
between these two form factors as ${F_0(p^2)}/{F_1 (p^2)}=\!
1-{p^2}/{m_B^2}$\cite{ch}. So we need only one of the two form
factors, say, $F_1$ to describe the decay amplitude just like that
in Eq.(\ref{a}).

Since our main task in this letter is to explore the infrared
unsafety due to the vertex corrections, we only give the explicit
results below for the infrared divergence terms in $f_I$ which are
regularized within the gluon mass scheme. We also present the
leading-twist results for $f_{II}$. Here $f_I$ and $f_{II}$
correspond respectively to the $T^I, T^{II}$ terms in
Eq.(\ref{qcdf}).

For ${}^3\!P_1$ charmonium state $\chi_{c1}$, there is no infrared
divergence in $B \to\! \chi_{c1}K$ decay, and we will give the
reasoning shortly (results of this decay within the light-cone
wave function description can be found in Ref.\cite{song2}).

For ${}^3\!P_0$ charmonium state $\chi_{c0}$,
 \bqa\label{fI1}
f_I\!=\!\frac{8 m_b z(1-z+\ln{z})}{(1-z)\sqrt{3
 z}}\ln{(\frac{\lambda^2}{m_b^2})}+finite\,\,terms,
   \eqa
 \bqa\label{fII1}
f_{II}=\!&\!&\!\frac{2}{ m_b (1-z)\sqrt{3 z}} \int_0^1 d\xi
\frac{\phi_B
(\xi)}{\xi} \int_0^1 dy \frac{\phi_K (y)}{{\bar y}^2}\nonumber\\
&&\times[-2 z + (1 - z)\bar y ].
 \eqa

 For ${}^3\!P_2$ charmonium state $\chi_{c2}$,
  \bqa\label{fI2}
   f_I\!\!=\!\!\frac{32 \epsilon_{\mu\nu}^{\ast J} p_b^\mu p_b^\nu
\!\sqrt{z^3}(\!1\!-\!z\!+\!\ln{z}\!)}{m_b
(1-\!z)^3}\!\ln{\!(\!\frac{\lambda^2}{m_b^2})}\!\!+\!\!finite\,\,terms,
\eqa
 \bqa\label{fII2}
f_{II}=\!&\!&\!\frac{8 \epsilon_{\mu\nu}^{\ast J} p_b^\mu
p_b^\nu\sqrt{z}}{ m_b^3 (1-z)^3} \int_0^1\!d\xi \frac{\phi_B
(\xi)}{\xi} \int_0^1\! dy \frac{\phi_K (y)}{{\bar y}^2} \nonumber\\
&&\times[z + (1 - z)\bar y ].
 \eqa

Here $z\!=\!M^2/m_B^2 \approx 4m_c^2/m_b^2$, $\xi$ is the momentum
fraction of the spectator quark in the $B$ meson and $y=(1-\bar
y)$ is the momentum fraction of the $s$ quark inside the $K$
meson, $\phi_B$ and $\phi_K$ are the light-cone wave functions for
the $B$ and $K$ mesons respectively\cite{BBNS}. Here $\lambda$ is
the gluon mass introduced to regularize the infrared divergences
in the vertex corrections. We have simplified the results for
$f_{II}$ by noting that $\xi\sim\mathcal{O}(\Lambda_{\rm
QCD}/m_b)\rightarrow 0$.

The integrals for $f_{II}$ in Eqs.(\ref{fII1}),(\ref{fII2}) will
give logarithmic divergences when the asymptotic form $\phi_K
(y)=6 y (1-y)$ for the kaon twist-2 light-cone distribution
amplitude as well as the same parameterization for the $\xi$
integration as that in Ref.\cite{BBNS} are used. It is also
worthwhile to emphasize that the infrared singularities in $f_I$
are more serious than the end-point singularities in $f_{II}$.
This is because, when one considers the effects of parton
transverse degrees of freedom for $B$ and $K$ mesons, end-point
singularities arising from spectator interactions can be
regularized \cite{kt} and logarithmic divergences can then be
removed. However, the infrared divergences due to nonfactorizable
vertex corrections still exist.

Since two large scales are involved, we can have several different
choices of the heavy quark limit. In the above calculations, we
have chosen the heavy quark limit as $m_b\to\! \infty$ with
$m_c/m_b$ fixed. In this limit, the infrared divergences in $f_I$
as well as the logarithmic divergences in $f_{II}$, will break
down QCD factorization. Another choice of the heavy quark limit is
that $m_b\to\! \infty$ with $m_c$ fixed. Then all the divergences
mentioned above are power corrections and should be dropped out,
so QCD factorization still holds in this limit. Physically, the
latter case is equivalent to the limit of zero charm quark mass in
which charmonium is regarded as a light meson. Obviously, the
first choice of the heavy quark limit with $m_c/m_b$ fixed is more
interesting in phenomenological analysis, and is also usually used
in theoretical studies\cite{BBNS,beneke}, where it is expected
that QCD factorization should apply to $B$ meson exclusive decays
into charmonium in the limit $m_c\to\!\infty$ with corrections of
order $\Lambda_{\rm QCD}/(m_c\alpha_s)\sim\! 1$. Our result shows
that this expectation holds for decays to S-wave charmonia but not
for P-wave charmonia we studied above where the vertex infrared
divergence will break down factorization at order of $\Lambda_{\rm
QCD}/(m_c\alpha_s)$.

It is worthwhile to discuss the origin of the  infrared
divergences and the differences between S-wave and P-wave
charmonia as well as the differences among P-wave charmonium
states.  We write momenta for $c\bar c$ as $p_c=p/2+q$, $p_{\bar
c}=p/2-q$. The gluon coupling to the $c\bar c$ pair showing in
Fig.\ref{fvs} is given by
 \bqa
J_\nu=\frac{\gamma_\nu(\pslash p_c+\pslash k
+m_c)\Gamma}{(p_c+k)^2-m_c^2}-\frac{\Gamma(\pslash p_{\bar
c}+\pslash k -m_c)\gamma_\nu}{(p_{\bar c}+k)^2-m_c^2},
 \eqa
where $\Gamma$ denotes for the weak decay vertex. When $k$ is
soft, the coupling in the nonrelativistic expansion in terms of
$q\cdot k/p\cdot k$ can be simplified to
  \bqa J_\nu
\label{IR} \approx 4\Gamma\Bigl[\frac{q_\nu}{p\cdot
k}-\frac{(q\cdot k) p_\nu}{(p\cdot k)^2}\Bigr],
 \eqa
where we used the on-shell conditions for $c$ and $\bar c$ quarks.
For S-wave charmonium states such as $J/\psi$, since the wave
functions depend on $q^2$, $q$ in Eq.(\ref{IR}) makes no
contribution to the lowest order. So there is no infrared
divergence for S-wave charmonium states. For $\chi_{c1}$, because
the orbital and spin momenta couplings are in an anti-symmetric
form (see Eq.(\ref{pol})), when combined with the $B\to K$ form
factors the infrared divergences arising from Eq.(\ref{IR}) are
totally cancelled out (note that this cancellation only
accidentally holds because of the specific form of $B\to K$ form
factors). However, for $\chi_{c0}$ and $\chi_{c2}$, because the
orbital and spin momenta couplings are in a symmetric form,
infrared divergences arising from Eq.(\ref{IR}) are not cancelled
out. This qualitative argument is confirmed by our explicit
calculations performed above.

It is well known that there are infrared divergences in the
\emph{inclusive} decay and production of P-wave charmonia, which
are related to the problems encountered here. In the \emph{
inclusive} processes, the infrared divergences in the
color-singlet P-wave $c \bar c$ state can be removed by including
contributions from the higher Fock states with color-octet $c \bar
c$ pair (say in S-wave) and soft gluon within the NRQCD
factorization framework (e.g. for B decay see \cite{bbyl}, for
annihilation hadronic decay see \cite{BBL2,huang}, and for decay
and production see \cite{petrelli,chen,ma}). However, to the best
of our understanding, the color-octet $c \bar c$ pair with
dynamical soft gluon can make contributions to the multi-body but
not two-body \emph{exclusive} decays. (Note that the soft gluon
emitted by the color-octet $c\bar c$ and subsequently reabsorbed
by the spectator light quark has already been considered in the
nonfactorizable spectator contribution discussed above (see
diagrams (e,f) in Fig.\ref{fvs}), where the soft gluon is related
to the end-point singularities of $B$ and $K$ light-cone wave
functions.) As a result, it seems to us that the infrared
divergences encountered in \emph{exclusive} processes involving
charmonia may raise a new question to the QCD factorization and
NRQCD factorization in B exclusive decays. Further studies are
needed to seek the solution to remove the infrared divergences.

In summary, we have studied the exclusive two-body decays of $B$
meson into the spin-triplet P-wave charmonium states
$\chi_{cJ}(J=\!0,1,2)$ and kaon within QCD factorization and NRQCD
by adopting the color-singlet non-relativistic wave function for
charmonium. We find that for these decays (except
$B\to\chi_{c1}K$), there are infrared divergences arising from
nonfactorizable vertex corrections, which can not be removed by
including contributions from the color-octet $c \bar c$ with soft
gluon Fock states, and therefore will break down QCD
factorization. New considerations should be introduced to describe
$B$ meson exclusive decays to charmonium states.

\section*{Acknowledgements}
We would like to thank G.T. Bodwin, E. Braaten, J.P. Ma, and J.W.
Qiu for helpful comments and K.-Y. Liu for useful discussions.
This work was supported in part by the National Natural Science
Foundation of China, and the Education Ministry of China.

\end{document}